\documentclass[12pt]{article}
\usepackage{amsmath,amsfonts,epsfig,amssymb}
\usepackage{graphics}

\begin{document}
\begin{center}
{\Large \bf  On the interpretation of the cosmic-ray anisotropy at
  ultra-high energies}

\medskip
D.S.~Gorbunov$^{a}$, 
P.G.~Tinyakov$^{a,b}$, 
I.I.~Tkachev$^{a}$, 
S.V.~Troitsky$^{a}$.
\\
\medskip
$^a${\small
Institute for Nuclear Research of the Russian Academy of Sciences,\\
60th October Anniversary prospect, 7a, 117312, Moscow, Russia.}\\
$^b${\small Service de Physique
Th\'eorique, Universit\'e Libre de Bruxelles, \\
CP225, blv.~du Triomphe, B-1050, Bruxelles, Belgium}
\end{center}
\vspace{0.5cm}

\begin{abstract}
A natural interpretation of the correlation between nearby Active
Galactic Nuclei (AGN) and the highest-energy cosmic rays observed
recently by the Pierre Auger Collaboration is that the sources of the
cosmic rays are either AGN or other objects with a similar spatial
distribution (the ``AGN hypothesis''). We question this
interpretation. We calculate the expected distribution of the arrival
directions of cosmic rays under the AGN hypothesis and argue that it
is not supported by the data, one of manifestations of the discrepancy
being the deficit of events from the direction of the Virgo
supercluster.  We briefly discuss possible alternative explanations
including the origin of a significant part of the observed events from
Cen A.
\end{abstract}

%\PACS{98.70.Sa}

\vspace{0.5cm}

%\sloppy

\section{Introduction}

The Pierre Auger Collaboration reported a remarkable correlation
\cite{PA} between the arrival directions of ultra-high-energy cosmic
rays (UHECR) and positions of nearby Active Galactic Nuclei (AGN). The
correlation was found by scanning over the maximum angular separation,
the minimum event energy and the maximum AGN redshift, see
Refs.~\cite{TT,Tinyakov:2003bi,Finley:2003ur} for the details of the method. 
The best signal was
found at the angle of 3.1$^\circ$ for the cosmic-ray set consisting of
15 events with reconstructed energies $E>5.6\times 10^{19}$~eV and for
the set of 472 AGN obtained by imposing the cut on the redshift, $z
\le 0.018$, in the catalog~\cite{Veron}. The correlation was tested
with the independent set consisting of 13 events, with the parameters
fixed {\it a~priori} from the first data set. The probability that the
correlation has occurred by chance is $1.7\times 10^{-3}$ as derived
from the independent set. The conclusion was made that there exists
anisotropy of arrival directions which is consistent with the
hypothesis that most of the cosmic rays reaching the Earth in that energy
range are protons from nearby astrophysical sources, either AGN or
other objects with a similar spatial distribution. We refer to the
latter proposition as the ``AGN hypothesis'' in what follows.

Crucial ingredients of the AGN hypothesis are i)~nearly rectilinear
propagation of UHECR and ii)~a large number of sources distributed
similarly to AGN, i.e., tracing the distribution of matter in the
Universe~\cite{AGN-distrib}. The second point deserves a comment. One
may estimate the number of sources contributing to the flux from the
statistics of clustering at small angles~\cite{DTT}. Counting events
separated by less than 6$^\circ$ as a doublet and adopting the
(unrealistic) assumption of equal flux on the Earth from all sources,
Ref.~\cite{PA1} estimates the lower limit on the number of sources as
61; a more fair assumption of equal cosmic-ray luminosity of the
sources raises the estimate to 252. Account of possible distribution
of sources in luminosity would further increase the estimate
of their number. The AGN hypothesis thus implies the existence of at
least a few hundred sources in the nearby Universe.

In this paper we show that, given the data presented in
Refs.~\cite{PA,PA1}, the AGN hypothesis is unlikely, and that there exist
alternative interpretations of the observed correlation. We should
stress that we question neither the fact nor the derived significance
of the correlation. However, by itself, the observed correlation of
UHECR with AGN implies only that the space distribution of UHECR at
highest energies is {\it not uniform}. The question of what are actual
sources of UHECR requires further study.  Here we make an attempt in
this direction. A brief account of our results has been reported in
our comment~\cite{comment} to the Pierre Auger publication.

\section{Testing the AGN hypothesis}

Let us consider the AGN hypothesis and see what kind of spatial
distribution of cosmic-ray flux it predicts. This flux depends on
several factors\footnote{First two factors are not accounted for in the
method of positional correlations used in Refs.~\cite{PA,PA1}.}: distances
to AGN, their luminosities and their spatial
distribution. The distance effect is accounted for by weighting each
source with the $1/r^2$ suppression factor, $r$ being the distance to
the source. The distribution of AGN in the cosmic-ray 
luminosities is not known. However, for a large group of sources
in a given region of the sky (e.g., a rich cluster of galaxies) the effect of
this distribution averages away. Therefore, in the case when the 
comic-ray flux is associated with the large-scale structure as in the AGN
hypothesis, one may consider the sources as having the same 
luminosities equal to the average source luminosity, provided this
average source luminosity is space-independent.

The space distribution of sources (which, under the AGN hypothesis,
are assumed to trace visible matter) in the nearby Universe is very
inhomogeneous. The role of local inhomogeneities is enhanced by the
cosmic-ray attenuation that cuts off the (uniform) flux coming from
distant sources. The AGN hypothesis implies, therefore, strongly
anisotropic flux at highest energies, with major local structures such
as the Centaurus and Virgo superclusters providing sizeable
contributions \cite{EGmogila}.

\begin{figure}
\begin{center}
\includegraphics[width=0.95\textwidth]{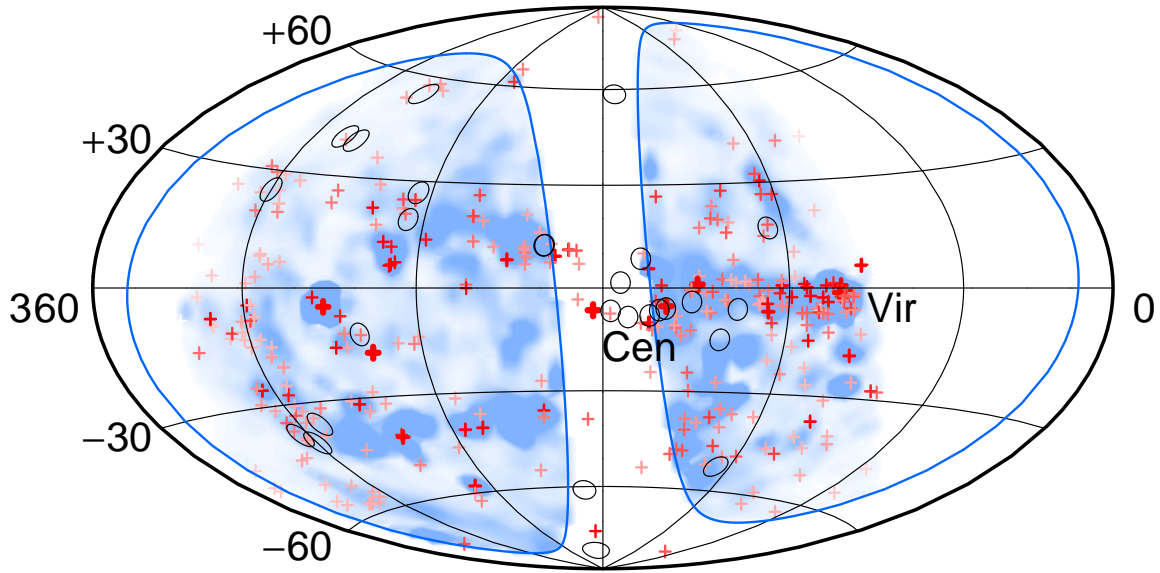}
\end{center}
\caption{
\label{fig:map}
Hammer projection of the celestial sphere in supergalactic
coordinates with crosses at the positions of nearby AGN from the
sample used in the correlation analysis of Ref.~\cite{PA}. The
color saturation of a given cross indicates the expected cosmic-ray flux
with the account of the exposure of the Pierre Auger Observatory (PAO)
and the $1/r^2$ suppression, $r$ being the distance to the source. Open
circles represent 27 highest-energy cosmic rays detected by PAO. Shading
shows the expected cosmic-ray flux from sources that follow the local
matter distribution (for details see Ref.~\cite{khrenov}), smoothed at the
angular scale of $3.1^\circ$ and convoluted with the PAO exposure 
(darker regions correspond to higher cosmic-ray flux). Blue lines cut out
the region with Galactic latitude $|b|< 15^\circ$ where the latter flux
cannot be determined because of incompleteness of the source catalog. The
positions of the Centaurus (Cen) and Virgo (Vir) superclusters are
indicated. }
\end{figure}

Fig.~\ref{fig:map}
shows the UHECR events used in the analysis of Ref.\cite{PA} together
with the expected flux of cosmic rays simulated assuming the AGN
hypothesis (see figure caption for notations). One may identify the
Virgo and Centaurus superclusters. The expected numbers of events from
these two structures are nearly equal. It is seen in
Fig.~\ref{fig:map} that there is a deficit of observed events from
Virgo as well as from other local structures, except the Centaurus
supercluster. This suggests that the AGN hypothesis proposed in
Ref.\cite{PA} may be disfavored by the data.

To quantify this statement we took the sample of AGN used in the
analysis of Ref.~\cite{PA}. According to the catalog
classification~\cite{Veron}, the sample consists of 457 AGN, 14
quasars and one probable BL Lac object, Cen~A. We removed from the
sample 3 objects with $z=0$ classified as stars in the database
\cite{NED}. We then calculated the space distribution of the expected
cosmic-ray flux, weighting each source with the factor $1/r^2$.

Fig.~2 shows the expected number of events within a given angular
distance from the center of the Virgo supercluster vs. the observed one.
\begin{figure}
\begin{center}
\includegraphics[width=0.95\textwidth]{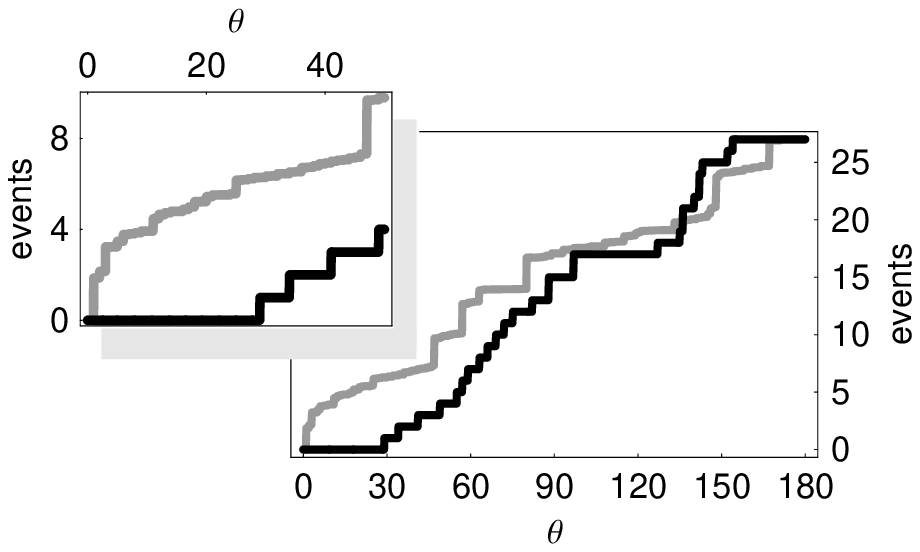}
\end{center}
\caption{
\label{fig:Vir}
Number of events in the circle of radius $\theta$ (in degrees) centered on
the Virgo cluster as determined in~\cite{NED}: gray, expected number of
events assuming the AGN hypothesis; black, events actually observed,
Ref.~\cite{PA}. The side panel zooms on the region around Virgo.}
\end{figure}
The observed and expected distributions
of events are inconsistent.  According to the Kuiper test, the probability
that the observed and simulated events are drawn from the same
distribution is $2\times 10^{-4}$. The main origin of inconsistency is
clear: out of 27 events, $\sim 6$ are expected to come from Virgo under the
AGN hypothesis while none is observed. The probability of this is
$10^{-3}$, in agreement with the Kuiper test.

Similar results are obtained in tests which do not use the Virgo
supercluster as a reference point. Comparing the Galactic longitudes
of observed and expected cosmic rays we find that the probability that
the two samples are drawn from the same distribution is 2\% according
to the Kuiper test, while for the Galactic latitude the corresponding
probability is $10^{-4}$. Analogous tests for the Supergalactic
longitude and latitude give the probabilities of $7\%$ and $10^{-4}$,
respectively.

The results do not change qualitatively when the catalog of AGN is
replaced by a complete sample of galaxies within the sphere of radius
270~Mpc \cite{khrenov}. Thus, our conclusions do not depend crucially
on specific properties of the AGN catalog used in Refs.~\cite{PA,PA1}
(in particular on its incompleteness), and apply equally to any class
of sources as long as they are distributed similarly to the visible
matter.  Given the strongest disagreement at the level of $10^{-4}$
and the number of various tests performed, we conclude that the AGN
hypothesis is disfavored at the confidence level of at least~99\%.

To demonstrate that our conclusions are insensitive to the assumption
of equal cosmic-ray luminosity of sources, we performed the same tests
assuming that at the Earth 
the cosmic-ray flux produced by an AGN is proportional to
its flux in visible light (derived from $V$-band magnitudes listed in
the catalog \cite{Veron}). No additional $1/r^2$ suppression is needed
in this case. This change in assumptions did not change the result
qualitatively. This is expected, as the number of AGN in the direction
of the Centaurus and Virgo superclusters is large (several dozens),
and the effect of the distribution in luminosity averages
away.

\section{Alternative explanations: Cen A?}

If, as it is suggested by the above arguments, the highest-energy
cosmic rays observed by the Pierre Auger Observatory do not come from
sources that follow the local matter distribution, how can one explain
the observed correlations with AGN? One possible explanation could be
the existence of a nearby bright source which happens to be in the
direction to the Centaurus supercluster where the density of
background AGN is larger than in average.  Events produced by such a
source and deflected by magnetic fields would overlap with the
background AGN more often than the uniformly distributed events, thus
creating a correlation signal with AGN. This spurious signal would
increase with the accumulation of statistics. Contrary to the AGN
hypothesis, this explanation would imply one or at most a few sources
in the nearby Universe and relatively large deflections, due to either
strong magnetic fields or to the presence of heavy nuclei in the
flux. Interestingly, in that case the properties of the highest-energy
cosmic rays may appear different in the Southern and Northern
hemispheres.

One plausible candidate for the source projected onto the Centaurus
supercluster is Cen~A~\cite{comment}. We would like to emphasize that
Cen A is not located within the Centaurus supercluster: Cen A is at
the distance of 3.5 Mpc from us, while the supercluster is at 45 Mpc.
Cen A is a special object, different from the majority of AGN in the
sample used in Ref. \cite{PA}. It is the closest Fanaroff-Riley type I
(FR I) radio galaxy. FR I galaxies constitute the parent population
for BL Lac type objects, i.e., it is believed that if the angle
between the jet of FR I galaxy and the line of sight is small, such an
object appears for an observer as a BL Lac type object.  Cen A
possesses the spectral energy distribution typical for BL Lacs
\cite{Chiaberge:2001ek}.  However, the value of the viewing angle
$\theta$ of its jet is still controversial. For the parsec-scale jet,
Ref. \cite{CenA_angle1} found $\theta \sim 50^\circ - 80^\circ$,
whereas Ref. \cite{CenA_angle2} found indications for $\theta \sim
15^\circ$ for the 100 pc scale jet. Cen A is often called a hidden or
misaligned BL Lac, and it is the only object classified as a BL Lac in
the catalog \cite{Veron} which enters the sample of Ref. \cite{PA}. We
should mention also that there have been correlations found between BL
Lacs and cosmic rays of lower energies in the Northern hemisphere
\cite{TT,MFmodel1,Gorbunov:2002hk,Gorbunov:2004bs,Abbasi:2005qy},
which are not supported, however, in the Southern hemisphere
\cite{Harari:2007up}.

Being the nearest FR~I radio galaxy, Cen A was suggested long ago as a
source of some~\cite{Cavallo,Ancho:Cen} or even most
\cite{Farrar,Weiler} of the observed cosmic rays of extreme
energies. (See, however, Ref.~\cite{Lemoin} where a different
interesting explanation of the Auger anisotropy is proposed and the
acceleration power of Cen~A is questioned.)  After the deficit of
events from Virgo in the Auger sample was argued to be a
problem~\cite{comment}, the interest to the conjecture has been
revived~\cite{Wolfendale,Fargion}.  Here, we support the plausibility
of the Cen-A scenario by the following observations.

Firstly, the study of the chemical composition of the primary UHECR
flux with the fluorescent detector of the Pierre Auger
observatory~\cite{PAO-comp} suggests that a significant fraction of
primary particles are heavy, or at least intermediate-mass nuclei. The
same conclusion has been recently made from the analysis of precise
Yakutsk muon data~\cite{Yak-comp}.  This agrees with the expectation
that heavier nuclei are generally accelerated to higher energies than
protons, which may be crucial for Cen A \cite{Neronov:2007mh} with its
weak acceleration power. However, heavy composition contradicts the
AGN hypothesis which assumes that most of the cosmic rays of highest
energies are protons since the expected deflections of heavy nuclei in
the Galactic magnetic field (GMF) are not compatible with the angular
scale of the observed correlation. Indeed, the deflections in the
$O(\mu G)$ regular component of the Galactic magnetic field can reach
a few tens of degrees for heavy nuclei. Additionally, one expects
deflections in the random component of the magnetic field which may be
comparable to the deflection in the regular component
\cite{TT:GMFrandom}.

Secondly, the outer radio lobes of Cen~A extend for $\sim 10^\circ$ in
the North--South direction (see, e.g.,\ Ref.~\cite{CenA:lobes}). Since
the lobes and hot spots are probable acceleration
sites~\cite{Biermann}, a number of the observed events can be
associated with Cen~A even without large deflections (i.e. assuming
that these events are protons).

\begin{figure}
\begin{center}
\includegraphics [width=0.95\textwidth]{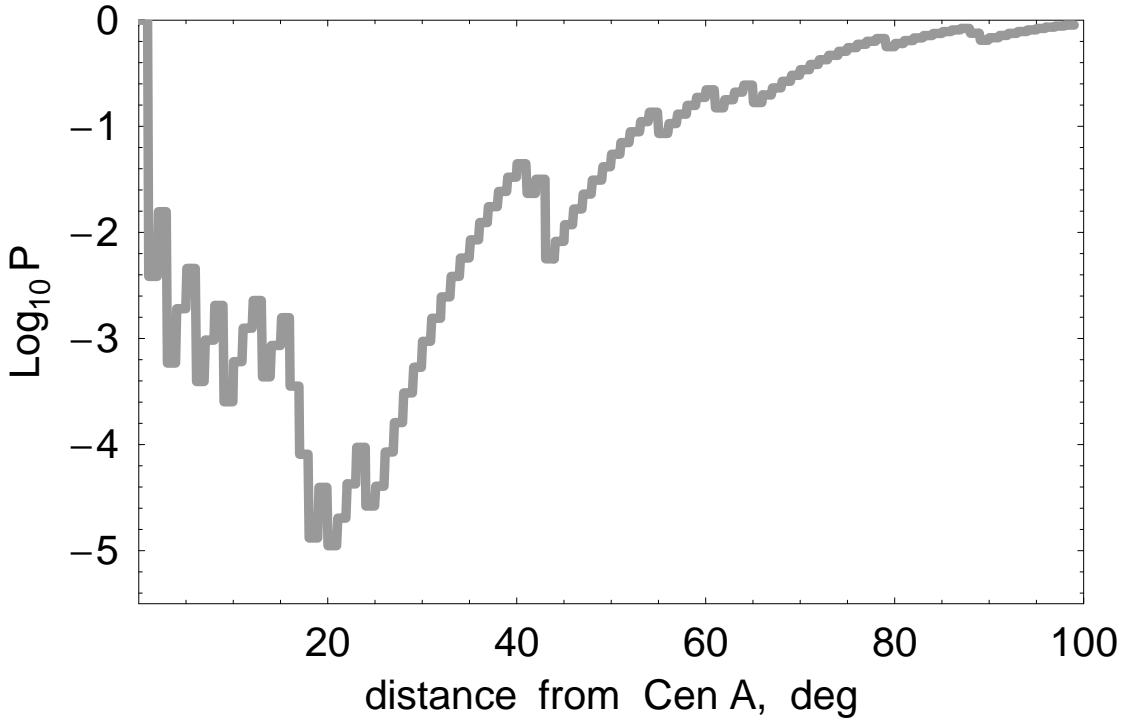}
\end{center}
\caption{
\label{fig:corr}
The Monte-Carlo probability $P$ to have the observed number
of events in a circle of a given radius around Cen~A on the celestial
sphere, out of the 27 events of the Auger sample. The values of $P$
are indicative only since their calculation accounts neither for the
statistical penalty associated with the choice of angular scale
nor for the bias in the sample.}
\end{figure}

Thirdly, there exists a correlation between the arrival directions of
UHECR and the position of Cen~A. The correlation test is similar to
that performed in Refs.~\cite{PA,PA1} except the catalog of candidate
sources now consists of only one object, Cen~A.  One finds (see
Fig.~\ref{fig:corr}) that, on average, for the isotropic
distribution 1.5 events out of 27 are expected to fall within the
circle of 20$^\circ$ around Cen~A, while 9 are observed. Out of $10^6$
Monte-Carlo tries this happened 11 times.

We see that the expected magnetic deflections of heavy nuclei and the
size of the outer radio lobes are in a good agreement with the angular
scale of the correlation signal around Cen~A, thus supporting the
hypothesis that Cen~A may be a source of a sizable part of the UHECR
events observed by PAO. A quantitative test of this or of any other
conjecture requires the full PAO data. The published subset used in
this analysis was obtained by tuning the cut on the minimal event
energy in such a way as to maximize the AGN correlation signal, and is
therefore biased\footnote{The sample of events used in Ref.~\cite{PA}
consisted of 81 events with reconstructed energies above 40 EeV and
zenith angles smaller than 60 degrees of which 27 remained after
tuning the cuts to maximize the effect reported there. Unfortunately
only this part (one third) of the used sample was made
public~\cite{PA1}.}.

\section{Conclusions}

To summarize, the data presented in Refs.~\cite{PA,PA1} disfavor the
hypothesis that most of the highest-energy cosmic rays come from a
large number of nearby astrophysical sources, either AGN or other
objects with a similar spatial distribution. Further doubts on this
interpretation arise from the absence of the similar correlation in
the Northern hemisphere, as one can find from the published AGASA data
and as it has been recently found by the HiRes
collaboration~\cite{Abbasi:2008md} (see, however, Ref.~\cite{YakAGN} for
a possible signal in the Yakutsk data). The alternative explanation of
the observed correlations could be, e.g., the existence of a bright
source in the direction of the Centaurus supercluster, Cen A being a
possible candidate. Both this and other interpretations could be
tested with larger unbiased data sets.

We are indebted to Valery Rubakov and Mikhail Shaposhnikov for
valuable comments. This work was supported in part by the grants of
the President of the Russian Federation NS-7293.2006.2 (government
contract 02.445.11.7370) (DG, IT and ST) and MK-1957.2008.2 (DG), by
the RFBR grant 07-02-00820, by the Russian Science Support Foundation 
(DG) and by the Belspo:IAP-VI/11, IISN and FNRS grants
(PT). We thank EPFL (DG), ULB (DG and ST) and CERN (ST) for
hospitality. Numerical simulations have been performed at the computer
cluster of the Theoretical Division of INR RAS. This work has made use
of the NASA/IPAC Extragalactic Database~\cite{NED} operated by the Jet
Propulsion Laboratory, California Institute of Technology, under
contract with NASA.

%%%%%%%%%%%%%%%%%%%%%%%%%%%%%%%%%%%%%%%%%%%%%%%%%%%%%%%%%%%%%%%%%%%%%%%%%%%%

\end{document}